\documentclass[12pt]{article}

\usepackage{amsmath,amssymb}
\usepackage{graphicx}

\makeatletter

\@addtoreset{equation}{section}
\makeatother

\newcommand{\be}{\begin{equation}}
\newcommand{\ee}{\end{equation}}
\newcommand{\bea}{\begin{eqnarray}}
\newcommand{\eea}{\end{eqnarray}}
\newcommand{\beann}{\begin{eqnarray*}}
\newcommand{\eeann}{\end{eqnarray*}}

\newcommand{\ba}{\begin{array}}
\newcommand{\ea}{\end{array}}

\newcommand{\N}{{\cal N}}

\def\XXint#1#2#3{{\setbox0=\hbox{$#1{#2#3}{\int}$} 
\vcenter{\hbox{$#2#3$}}\kern-.5\wd0}}

\begin{document}

\setlength{\oddsidemargin}{0cm}
\setlength{\baselineskip}{7mm}

\begin{titlepage}
\renewcommand{\thefootnote}{\fnsymbol{footnote}}


~~\\

\vspace*{0cm}
    \begin{Large}
    \begin{bf}
       \begin{center}
         { Kerr/CFT correspondence \\
and \\
five-dimensional BMPV black holes}
      
       \end{center}
    \end{bf}   
    \end{Large}
\vspace{1cm}

   \begin{large}
\begin{center}
{ \sc Hiroshi Isono}$^2$\footnote    
{e-mail address : 
isono@ntu.phys.edu.tw}, 
{ \sc Ta-Sheng Tai}$^1$\footnote    
{e-mail address : 
tasheng@riken.jp} and 
{ \sc Wen-Yu Wen}$^{2,3}$\footnote    
{e-mail address : 
steve.wen@gmail.com}
\end{center}
\end{large}

      \vspace{1cm}

\begin{center}
 1~~{\it   Theoretical Physics Laboratory, RIKEN,
                    Wako, Saitama 351-0198, JAPAN}
\end{center}
\begin{center}
     2~~{\it   Department of Physics and Center for Theoretical
      Sciences, National Taiwan University, Taipei 106, TAIWAN}
\end{center}   
\begin{center}
     3~~{\it   Leung Center for Cosmology and Particle Astrophysics, 
National Taiwan University, Taipei 106, TAIWAN}
\end{center}

\vspace{1cm}

\begin{abstract}
\noindent
We apply a recently proposed Kerr/CFT correspondence to extremal 
supersymmetric five-dimensional charged spinning black holes, constructed by 
Breckenridge, Myers, Peet and Vafa. By computing the central charge of the dual CFT and Frolov-Thorne temperature, Cardy's formula succeeds in reproducing  
Bekenstein-Hawking area law. 
\end{abstract}
\vfill 

\end{titlepage}
\vfil\eject

\setcounter{footnote}{0}

\section{Introduction}
Considerable progress in deriving 
black hole entropy statistically  
has been made by resorting to state counting approaches. 
Among them, while Cardy's formula in conformal field theory (CFT) 
plays an indispensable role, this can be better understood 
in the context of AdS/CFT correspondence via string compactification and wrapped branes \cite{Strominger:1996sh}. 
For example, a 2D ${\cal{N}}=(0,4)$ CFT living on an M5-brane wrapping 
spatially $S^1 \times P_4$ $(P_4 \subset CY_3)$ was shown to be dual to a 4D BPS 
black hole formed by a Type IIA D0-D2-D4 system which 
has an attractor geometry near its horizon \cite{Maldacena:1997de}. The entropy in terms of Cardy's formula 
\begin{align}
\begin{split}
{2\pi \sqrt{\frac{c_L L_0}{6}}}
\label{CA} 
\end{split}
\end{align}
agrees with Bekenstein-Hawking area law.  
Here $c_L$ denotes the central charge and $L_0$ is the eigenvalue
of left-moving Virasoro zero mode.

On the other hand, an alternative pioneered much earlier by 
Brown and Henneaux \cite{Brown:1986nw} is to take into account the 
asymptotic symmetry preserved at the boundary under 
suitable boundary conditions. 
They dealt with $AdS_3$ with $SL(2,R)_L \otimes SL(2,R)_R$ isometry which a 3D BTZ black hole asymptotically approaches. 
There, two copies of Virasoro algebra%
\footnote{See also 
\cite{Carlip:1998wz,Solodukhin:1998tc,Park:2001zn} for the 
appearance of Virasoro algebra in generic black holes.} 
emerge as a result of infinitely many 
Fourier modes of 
the boundary diffeomorphism $\xi^{\mu}(x) \partial_{\mu}$. 
The central term arising from 
commutators of Virasoro generators was later used to reproduce the macroscopic 
entropy $S_{BTZ}={2\pi \sqrt{\frac{c_L L_0}{6}}}+ {2\pi \sqrt{\frac{c_R \tilde{L}_0}{6}}}$ by 
Strominger \cite{Strominger:1997eq}.

In much the same spirit of Brown-Henneaux,
chiral auxiliary 2D CFTs dual to 4D extremal Kerr black holes 
have recently been proposed by Strominger et al. \cite{Guica:2008mu}. 
In their paper and a series of related works \cite{Hotta:2008xt,Lu:2008jk,Azeyanagi:2008kb,Hartman:2008pb,Nakayama:2008kg,Chow:2008dp}, 
on the black hole near-horizon geometry certain 
crucial boundary conditions are imposed such that 
the asymptotic symmetry group (ASG) preserving it includes 
ultimately two kinds of generators, i.e. 
\begin{align}
\begin{split}
K^t &= \partial_t ,\\
K^{\phi} &=\epsilon(\phi)\partial_{\phi}-r \epsilon'(\phi)\partial_r,
\label{K}
\end{split}
\end{align}
where $\phi$ denotes some angular coordinate and $r$ stands for the radial direction. 
Decomposing an arbitrary periodic $\epsilon(\phi)$ into infinitely many Fourier modes labeled by $n$, 
one may identify $K^{\phi}_n$ with 
the generator $L_n$ of Virasoro algebra. 
Consequently, the central charge $c$ can be determined completely 
from the near-horizon metric and \eqref{K} 
owing to techniques developed in literatures \cite{Barnich:2001jy,Barnich:2007bf}. 
Quite remarkably, by further introducing Frolov-Thorne temperature $T_{FT}$ associated with 
$\phi$ \cite{Frolov:1989jh}, 
Cardy's formula
\begin{align}
\begin{split}
S=\frac{\pi^2}{3} c T_{FT}
\label{C}
\end{split}
\end{align}
reproduces the macroscopic entropy perfectly. 
\eqref{C} can be regarded as a Legendre transformed version of 
\eqref{CA} with an effective $T_{FT}$.

In this article, 
we apply the above procedure as well as \eqref{C} to 
a well-known 5D extremal supersymmetric charged spinning black hole 
constructed by Breckenridge, Myers, Peet and Vafa (BMPV) \cite{Breckenridge:1996is}. 
Unlike 4D Kerr-Newman black holes, this solution still exhibits unbroken supersymmetry 
even extremality is satisfied.  
As BTZ black holes mentioned above, 
the microscopic origin of 
BMPV entropy first roots in its D-brane realization%
\footnote{It can also be realized via an M-theory lift 
of a D0-D2-D6 system wrapped on $CY_3$, see 
\cite{Yin:2006mb} and references therein.}. 
Nevertheless, the degeneracy counting that we derive below 
will rely thoroughly on Virasoro algebra stemming from ASG 
and the corresponding Frolov-Thorne temperature.

In section $2$, we introduce briefly the BMPV black hole, a solution to the equation of motion of  
5D Einstein-Maxwell-Chern-Simons gravity and present its conserved charges. 
In section $3$, we carry out the computation of its dual chiral CFT central charge and Frolov-Thorne temperature. 
By making use of Cardy's formula, 
perfect agreement with 
Bekenstein-Hawking area law is found. 
Finally, we conclude with some comments in section 4. 


\section{BMPV black hole and near-horizon geometry}

\subsection{BMPV black hole}
As shown in \cite{Kallosh:1996vy}, the BMPV solution can be 
embedded in 5D ${\cal{N}}=2$ supergravity and is charged under a graviphoton. 
The metric reads (Planck length $l_5=1$)
\begin{eqnarray}\label{BMPV}
&{}&
ds^2=
-\left(1-\frac{\mu}{r^2}\right)^2 dt^2
+\frac{dr^2}{\left(1-\frac{\mu}{r^2}\right)^2}
-\frac{\mu a}{r^2}\left(1-\frac{\mu}{r^2}\right)\sigma_3dt
-\frac{\mu^2a^2}{4r^4}\sigma_3^2
+\frac{r^2}{4}d\Omega_3^2, \nonumber\\
&{}&
\sigma_3=d\phi+\cos\theta d\psi,~~~~
d\Omega_3^2=d\theta^2+\sin^2\theta d\psi^2+\sigma_3^2 , \nonumber\\
&{}&
\theta \in [0,\pi], ~~~~\phi \in  [0,2\pi],~~~~\psi \in [0,4\pi] 
\end{eqnarray}
with gauge potentials 
\begin{eqnarray}
&{}& 
A=B(r)dt+C(r)\sigma_3, \nonumber\\
&{}&
B(r)=\frac{\sqrt{3}\mu}{2r^2},~~~~C(r)=-\frac{\sqrt{3}\mu a}{4r^2} 
\end{eqnarray}
and a constant dilaton field. $d\Omega_3^2$ is the line element on $S^3$. 

The conserved energy, angular momentum and graviphoton charge for the BMPV black hole are as follows: 
\begin{eqnarray}
M=\frac{3}{4}\pi\mu,~~~~
J=\frac{1}{4}\pi a\mu,~~~~
Q=\frac{\sqrt{3}}{2}\pi\mu,
\end{eqnarray}
which satisfy the first law of black hole thermodynamics 
($\Omega_{\psi}=0$)
\begin{equation}
dM = T_{H}dS + \Omega_{\phi} dJ + \Phi dQ.
\end{equation}
Due to extremality, Hawking temperature $T_{H}$ and 
angular velocity $\Omega_{\phi}$ are zero, while the chemical potential $\Phi$ is equal to $B(\sqrt{\mu})$. 
A tricky point is that the ratio $\frac{T_{H}}{\Omega_{\phi}}$ is 
definitely finite at $r=\sqrt{\mu}$. We will use this fact 
later in section $3.3$.

\subsection{Near-horizon geometry}
By taking near-horizon limit: 
$r=\sqrt{\mu}(1+ \frac{\lambda}{2} \hat{r})$ and $t=\frac{\sqrt{\mu}}{2\lambda} \hat{t}$ with $\lambda\rightarrow 0$, 
the BMPV metric (\ref{BMPV}) becomes 
\begin{eqnarray}
ds^2 &=& \frac{\mu}{4}(-\hat{r}^2 d\hat{t}^2 + \frac{d\hat{r}^2}{\hat{r}^2})-\frac{a\sqrt{\mu}}{2}\hat{r}(d\phi+\cos{\theta}d\psi)d\hat{t}\nonumber\\
&&+\frac{\mu-a^2}{4}(d\phi+\cos{\theta}d\psi)^2+\frac{\mu}{4}(d\theta^2+\sin^2{\theta}d\psi^2).
\label{BMPV2}
\end{eqnarray}
It is seen that \eqref{BMPV2} 
possesses 
a structure of $AdS_2$ (in Poincar\'{e} patch) fibered over $S^3$.
This strongly suggests at this 
limit the existence of a dual 2D chiral CFT 
whose central charge will be obtained thereof.  
Before proceeding to the dual CFT computation, we note 
that 
the horizon area is equal to 
\begin{align}
\begin{split}
A_{horizon}={2\pi^2}\mu\sqrt{\mu-a^2}
\label{H1}
\end{split}
\end{align} 
and thus the macroscopic Bekenstein-Hawking entropy is given 
by ($G_5=1$) 
\begin{align}
\begin{split}
S_{macro}=\frac{A_{horizon}}{4}=\frac{\pi^2}{2}\mu\sqrt{\mu-a^2}.
\label{H2}
\end{split}
\end{align}

\section{Entropy from chiral CFT} 

\subsection{Boundary condition and asymptotic symmetry group}

Following the work \cite{Brown:1986nw}, 
we have to impose carefully 
boundary conditions on the asymptotic variation of the metric \eqref{BMPV2} and single out the desired ASG.

Let $h_{\mu\nu}$ be the perturbation around the near-horizon metric (\ref{BMPV2}). 
We choose the following boundary condition:
\begin{align}
\left(
\begin{array}{ccccc}
h_{tt}=\mathcal{O}(r^2)
&h_{tr}=\mathcal{O}(\frac{1}{r^2})
&h_{t\theta}=\mathcal{O}(\frac{1}{r})
&h_{t\phi}=\mathcal{O}(1)
&h_{t\psi}=\mathcal{O}(r)\\
h_{rt}=h_{tr}
&h_{rr}=\mathcal{O}(\frac{1}{r^3})
&h_{r\theta}=\mathcal{O}(\frac{1}{r^2})
&h_{r\phi}=\mathcal{O}(\frac{1}{r})
&h_{r\psi}=\mathcal{O}(\frac{1}{r^2}) \\
h_{\theta t}=h_{t\theta}
&h_{\theta r}=h_{r\theta}
&h_{\theta\theta}=\mathcal{O}(\frac{1}{r})
&h_{\theta\phi}=\mathcal{O}(\frac{1}{r})
&h_{\theta \psi} = \mathcal{O}(\frac{1}{r}) \\
h_{\phi t}=h_{t\phi}
&h_{\phi r} =h_{r\phi}
&h_{\phi\theta} =h_{\theta\phi}
&h_{\phi\phi}=\mathcal{O}(1)
&h_{\phi \psi}=\mathcal{O}(1)\\
h_{\psi t}=h_{t\psi}
&h_{\psi r}= h_{r\psi}
&h_{\psi\theta}=h_{\theta \psi}
&h_{\psi\phi} =h_{\phi \psi}
&h_{\psi\psi}=\mathcal{O}(\frac{1}{r})
\end{array}
\right).
\label{boundary_condition}
\end{align}
The most general diffeomorphism which respects 
the above boundary condition reads
\begin{equation}
\zeta =
\Bigl[C+\mathcal{O}\bigl(\frac{1}{r^3}\bigr)\Bigr]\partial_t
+[-r\epsilon'(\phi)+\mathcal{O}(1)]\partial_{r}
+\mathcal{O}\bigl(\frac{1}{r}\bigr)\partial_{\theta} +\mathcal{O}\bigl(\frac{1}{r^2}\bigr)\partial_{\psi}
+\Bigl[\epsilon(\phi)+\mathcal{O}\bigl(\frac{1}{r^2}\bigr)\Bigr]\partial_\phi,
\end{equation}
where $C$ is an arbitrary constant and
$\epsilon(\phi)$ is an arbitrary periodic function of $\phi$. 
We have dropped the hat over $t$ and $r$ for brevity. 
As a result, ASG here is simply generated by 
\begin{eqnarray}
&&\zeta^t=\partial_{t}, \label{time_killing}\nonumber\\
&&\zeta^{[1]}=\epsilon(\phi)\partial_\phi-r\epsilon'(\phi)\partial_r.
\end{eqnarray}

\subsection{Central charge}
We use the method developed in \cite{Barnich:2001jy,Barnich:2007bf} 
to compute the central charge of the dual chiral CFT.
Let us start with 
\begin{eqnarray}
&&\zeta^{[1]}_{(n)}=-e^{-in\phi}\partial_{\phi}-inre^{-in\phi}\partial_r,\nonumber\\
&&\zeta^{[2]}_{(n)}=-e^{-in\psi}\partial_{\psi}-inre^{-in\psi}\partial_r,
\label{gvec}
\end{eqnarray}
which correspond to Fourier modes of the periodic function $\epsilon$. Naively, commutators of $\zeta^{[i]}_{(n)}$'s 
constitute two copies of chiral Virasoro algebra without central terms. 
Nevertheless, the central extension $c^{(j)}$ $(j=1,2)$ can 
be given as follows: 
\begin{eqnarray}\label{c}
&{}&
\frac{1}{8\pi}\int_{\partial\Sigma}k_{\zeta_{(m)}^{[j]}}[{\mathcal{L}_{\zeta_{(n)}^{[j]}}}g, g]=
-\frac{i}{12}(m^3+\xi m)c^{(j)}\delta_{m+n,0},
\end{eqnarray}
where $\partial\Sigma$ is a spatial slice and 
$\mathcal{L}_{\zeta}$ denotes Lie derivative 
with respect to $\zeta$. The 3-form 
$k_{\zeta}$ is defined by
\begin{eqnarray}
&{}&
k_{\zeta}[h,g]=
\frac{1}{2}\left[\zeta_{\nu}D_{\mu}h-\zeta_{\nu}D_{\sigma}h_{\mu}^{~\sigma}
+ \zeta_{\sigma}D_{\nu}h_{\mu}^{~\sigma}
+\frac{1}{2}hD_{\nu}\zeta_{\mu}-h_{\nu}^{~\sigma}D_{\sigma}\zeta_{\mu}
+\frac{1}{2}h^{\sigma}_{~\nu}(D_{\mu}\zeta_{\sigma}+D_{\sigma}\zeta_{\mu})\right] \nonumber\\
&{}&~~~~~~~~~~~~*(dx^{\mu}\wedge dx^{\nu}),
\label{cc}
\end{eqnarray}  
where covariant derivatives and 
Einstein summation are performed with respect to $g_{\mu \nu}$. 
The coefficient $\xi$ in \eqref{c} is irrelevant because it can be absorbed by a shift of Virasoro zero mode.
Equipped with generators of ASG \eqref{gvec} and the near-horizon metric \eqref{BMPV2}, 
we have 
\begin{eqnarray}
(\mathcal{L}_{\zeta_n^{[1]}}g)_{tt} &=& \frac{ i}{2}\mu r^2ne^{-{ i}n\phi} \nonumber\\
(\mathcal{L}_{\zeta_n^{[1]}}g)_{t\psi} &=& \frac{ i}{4}anr\sqrt{\mu}\cos{\theta}\,e^{-{ i}n\phi} \nonumber\\
(\mathcal{L}_{\zeta_n^{[1]}}g)_{r\phi} &=& -\frac{1}{4}\frac{\mu n^2}{r}e^{-{ i}n\phi} \nonumber\\
(\mathcal{L}_{\zeta_n^{[1]}}g)_{\phi\phi} &=& -\frac{ i}{2}(a^2-\mu)ne^{-{ i}n\phi} \nonumber\\
(\mathcal{L}_{\zeta_n^{[1]}}g)_{\phi\psi} &=& -\frac{ i}{4}(a^2-\mu)n\cos\theta\,e^{-{ i}n\phi}
\end{eqnarray}
and 
\begin{eqnarray}
(\mathcal{L}_{\zeta_n^{[2]}}g)_{tt} &=& \frac{ i}{2}\mu r^2ne^{-{ i}n\psi} \nonumber\\
(\mathcal{L}_{\zeta_n^{[2]}}g)_{t\phi} &=& \frac{ i}{4}anr\sqrt{\mu}\cos{\theta}e^{-{ i}n\psi} \nonumber\\
(\mathcal{L}_{\zeta_n^{[2]}}g)_{r\psi} &=& -\frac{1}{4}\frac{\mu n^2}{r}e^{-{ i}n\psi} \nonumber\\
(\mathcal{L}_{\zeta_n^{[2]}}g)_{\phi\psi} &=& -\frac{i}{4}(a^2-\mu)n\cos\theta\,e^{-{ i}n\psi} \nonumber\\
(\mathcal{L}_{\zeta_n^{[2]}}g)_{\psi\psi} &=& -\frac{i}{2}(a^2\cos^2\theta-\mu)ne^{-{ i}n\phi}
\end{eqnarray}
Substituting these back to (\ref{cc}), 
we obtain
\begin{eqnarray}
c^{(1)}=3\pi a\mu,~~~~c^{(2)}=0.
\label{cen}
\end{eqnarray}
This result can be reasoned as below. 
Due to a different coordinate choice, 
in \cite{Kallosh:1996vy} the black hole has 
two equal but opposite spins $\pm J$. 
Here, by using Hopf fiber description of $S^3$, one of them turns into a spin $2J$ (associated with $\phi$ coordinate), while the other (associated with $\psi$ coordinate) vanishes.

\subsection{Frolov-Thorne temperature}
Let us determine the so-called Frolov-Thorne temperature. 
First, through equating eigen-modes near the horizon and 
elsewhere (hat is restored)
\begin{eqnarray}
e^{-i\omega t+im_{\phi}\phi+im_{\psi}\psi}=
e^{-im_R\hat{t}+im_{L\phi}\hat{\phi}+im_{L\psi}\hat{\psi}},
\end{eqnarray}
one has the relation between quantum numbers like 
\begin{eqnarray}\label{mmlr}
m_R=\frac{\omega\sqrt{\mu}}{2\lambda},~~
m_{L\phi}=m_{\phi},~~m_{L\psi}=m_{\psi}
\end{eqnarray}
for $t=\frac{\sqrt{\mu}}{2\lambda}\hat{t}$, 
$\phi=\hat{\phi}$ and $\psi=\hat{\psi}$.

Next, we rewrite Boltzmann factor as 
($\Omega_{\psi}=0$) 
\begin{eqnarray}
\exp\left(-\frac{\omega-\Omega_{\phi}m_{\phi}}{T_H}\right)=
\exp\left(-\frac{m_R}{T_R}-\frac{m_{L\phi}}{T_{\phi}}
\right).
\end{eqnarray}
$T_R$ and $T_{\phi}$ are Frolov-Thorne temperatures.
From \eqref{mmlr}, we get
\begin{eqnarray}
T_R=\frac{T_H\sqrt{\mu}}{2\lambda},~~~~
T_{\phi}=-\frac{T_H}{\Omega_{\phi}}.
\end{eqnarray}
Automatically, $T_R=0$ due to extremality and  
\begin{eqnarray}\label{FT}
T_{\phi}=-\lim_{r\rightarrow \sqrt{\mu}}\frac{T_H(r)}{\Omega_{\phi}(r)}=\frac{\sqrt{\mu-a^2}}{2\pi a}.
\label{limit}
\end{eqnarray}
As advertised, the ratio 
$\frac{T_H}{\Omega_{\phi}}$ remains non-vanishing 
at the horizon by carefully examining \eqref{limit}. We 
present this procedure in Appendix%
\footnote{We are grateful to Chiang-Mei Chen who pointed 
out that a missing factor 2 in version one 
may be attributable to 
our previous Frolov-Thorne temperature.}.

\subsection{Microscopic entropy}
Substituting \eqref{cen} and \eqref{FT} into Cardy's formula \eqref{C}, we obtain the microscopic entropy 
\begin{eqnarray}
S_{ micro}=\frac{1}{2}\pi^2\mu\sqrt{\mu-a^2}.
\end{eqnarray}
This agrees precisely with Bekenstein-Hawking entropy \eqref{H2}.

\section{Conclusion and comments} 

We have succeeded in reproducing the BMPV black hole entropy using Kerr/CFT correspondence. 
This non-trivial check suggests that counting 
entropy semi-classically by evaluating the dual CFT central charge and effective temperature is also applicable to 5D extremal 
supersymmetric charged spinning black holes. 
 
Since the central charge in BMPV cases is proportional to 
$J$ (spin), we expect
that a lifted solution in 6D similar to 
\cite{Hartman:2008pb} can be constructed in order to reproduce the
entropy in the degenerate limit $a \to 0$. 
Also, in ${\N=2}$ Type IIA string compactified on $CY_3$, 
BMPV black holes can be realized via an M-theory lift of D0-D2-D6 systems wrapped on $CY_3$ 
with brane charges $(q_0, q_A, 1)$\footnote
{$A$ is the index for the 2-cycle basis of $CY_3$.}. 
Due to one single D6-brane, 
the 5D black hole 
is located at the center of 
a Taub-NUT space and ${q_0}\propto J$ is associated with 
its spin over the $S^1$ bundle of Taub-NUT. 
In addition, $q_A$ is related to $Q$ by $q_A=\frac{3Q}{Y^A}$ 
where scalar fields $Y$'s in vector multiplets take their 
horizon values with normalization $1=D_{ABC} Y^A Y^B Y^c$ 
($D_{ABC}$: triple intersection number of $CY_3$). 
It will be interesting to check whether Kerr/CFT 
prescription works as well for generic 
configurations of IIA brane charges $(q_0,q_A,p^A,p^0)$. 


As another remark, the central charge in \eqref{cen} differs from 
what is microscopically derived in \cite{Breckenridge:1996is} where 
$c\propto \mu^2\sim Q^2$ for large $\mu$. 
This feature is not encountered in 3D BTZ cases because 
the central charge of Brown-Henneaux is exactly equal to 
that in the dual 2D ${\cal{N}}=(4,4)$ CFT \cite{Strominger:1997eq}.  
It remains interesting 
to understand the exotic 
Kerr/CFT correspondence by pursuing this discrepancy further.


\section*{Acknowledgements}
We are grateful to Yutaka Matsuo and 
Chiang-Mei Chen for valuable discussion. Also, 
we thank Sergey Solodukhin, 
Glenn Barnich and Mu-In Park for helpful comments. 
T.S.T is grateful to Xi Yin and Wei Li for an introduction to BPS black holes. 
We would like to thank Mathematica and RGTC package for facilitating our computation. 
T.S.T is supported in part by the postdoctoral program at RIKEN. 
H.I and W.Y.W are partially supported by Taiwan National 
Science Council under Grant No. 97-2119-M-002-001 and 97-2112-M-002-015-MY3. 

\appendix 

\section{Appendix} 
According to \cite{Wu:2007gg}, one is able to have a general 
black hole embedded in G\"{o}del universe parameterized by $(m,q,j,a)$, 
namely, 
\begin{eqnarray}
ds^2 = -f(r)dt^2 - 2g(r)\sigma_3 dt + 
h(r)\sigma_3^2 +
\frac{dr^2}{V(r)}
+\frac{r^2}{4} d\Omega_3^2, 
\end{eqnarray}
where 
\begin{eqnarray}
f(r)
&=& 1-\frac{2m}{r^2}+ \frac{q^2}{r^4},\\
V(r)
&=& 1-\frac{2m-8j(m+q)(a+2j(m+2q))}{r^2} \nonumber\\
&{}& +\frac{2(m-q)a^2+q^2(1-16ja-8j^2(m+3q))}{r^4}, \\
h(r)
&=& -j^2r^2(r^2+2m+6q)+3jqa+\frac{(m-q)a^2}{2r^2}-\frac{q^2a^2}{4r^4}, \\
g(r)
&=& jr^2+3jq+\frac{(2m-q)a}{2r^2}-\frac{q^2a}{2r^4}. 
\end{eqnarray}
Further, Hawking temperature $T_H(r_+)$ and 
the angular velocity $\Omega_{\phi}(r_+)$ 
($\Omega_{\psi}(r_+)=0$) at the outer horizon are defined via 
\begin{eqnarray}
T_H (r)&=&\frac{rV'(r)}{4\pi\sqrt{4h(r)+r^2}},
\nonumber\\
\Omega_{\phi}(r)&=&\frac{g(r)}{h(r)+\frac{r^2}{4}}.
\end{eqnarray}
The BMPV solution saturates $m=q$ with 
$j=\frac{\sqrt{2(m-q)}}{4(m+q)}$, 
and $r_+=r_- =\sqrt{m}$. 
Since both Hawking temperature and the angular velocity go to 
zero as $r_+ \rightarrow \sqrt{m}$, 
Frolov-Thorne temperature of BMPV 
\begin{eqnarray}
T_{\phi}=-\lim_{r\rightarrow \sqrt{m}}\frac{T_H(r)}{\Omega_{\phi}(r)}
\end{eqnarray}
should be calculated instead by    $\frac{T_H^{\prime}(r)}{\Omega_{\phi}^{\prime}(r)}$ 
according to l'Hopital's theorem. 

Adhering to the three steps in order: first setting $j=\frac{\sqrt{2(m-q)}}{4(m+q)}$, then taking $m=q$ 
followed by substituting $r=\sqrt{m}$ as the final step, 
one can show that 
\begin{eqnarray}
T_{\phi}=\frac{\sqrt{m-a^2}}{2\pi a}. 
\end{eqnarray}
This is nothing but \eqref{limit} by putting $m=\mu$.

\end{document}